# Water modeled as an intermediate element between carbon and silicon.

*Valeria Molinero and Emily B. Moore*

Department of Chemistry, University of Utah, 315 South 1400 East,
Salt Lake City, UT 84112, U.S.A.

e-mail: Valeria.Molinero@utah.edu

Corresponding Author:

Valeria Molinero, Department of Chemistry, University of Utah, 315 South 1400 East, Salt Lake City, UT 84112, U.S.A. Phone +1801-585-9618, fax +1801-581-4353, e-mail: Valeria.Molinero@utah.edu

**ABSTRACT**

Water and silicon are chemically dissimilar substances with common physical properties. Their liquids display a temperature of maximum density, increased diffusivity on compression, they form tetrahedral crystals and tetrahedral amorphous phases. The common feature to water, silicon and carbon is the formation of tetrahedrally coordinated units. We exploit these similarities to develop a coarse-grained model of water (mW) that is essentially an atom with tetrahedrality intermediate between carbon and silicon. mW mimics the hydrogen-bonded structure of water through the introduction of a nonbond angular dependent term that encourages tetrahedral configurations. The model departs from the prevailing paradigm in water modeling: the use of long-ranged forces (electrostatics) to produce short-ranged (hydrogen-bonded) structure. mW has only short-range interactions yet it reproduces the energetics, density and structure of liquid water, its anomalies and phase transitions with comparable or better accuracy than the most popular atomistic models of water, at less than 1% of the computational cost. We conclude that it is not the nature of the interactions but the connectivity of the molecules that determines the structural and thermodynamic behavior of water. The speedup in computing time provided by mW makes it particularly useful for the study of slow processes in deeply supercooled water, the mechanism of ice nucleation, wetting-drying transitions, and as a realistic water model for coarse-grained simulations of biomolecules and complex materials.

# I. Introduction

Computer simulations play an important role in understanding the significance of microscopic interactions in water properties. The first model of liquid water was proposed in 1933 by Bernal and Fowler: an ice-like disordered tetrahedral structure arising from the electrostatic interactions between close neighbors.[1] About hundred atomistic potentials of water have been developed since then. The apparent profligacy of atomistic potentials is not just a tribute to water's essential role in nature, but an admission of the difficulty in representing the complex physics of water with a simple, efficient to compute, model. Atomistic models used in molecular simulations use long-ranged forces (electrostatics) to produce short-ranged tetrahedral structure (hydrogen bonds). The most popular models of water -SPC[2], SPCE[3], TIP3P[4], TIP4P[4], TIP5P[5], and their polarizable cousins[6-9] -follow this modeling paradigm based on the electrostatic nature of the intermolecular interactions in real water.

In this article we address the question of what are the essential ingredients for a model to generate the thermodynamic, dynamic and structural anomalies of water,[10] while quantitatively reproducing water's experimental structure, energetics and phase behavior. Can a coarse-grained model without electrostatic interactions and hydrogen atoms reproduce the structure and phase behavior of water as accurately as all-atoms models?

The idea of developing a coarse-grained model of water, without hydrogen and electrostatics, is not new.[11-14] Here we make a distinction between coarse-grained and toy models of water: the former are parameterized to quantitatively reproduce some water properties, while the latter aims to qualitatively capture water's anomalous behavior without attempting to reproduce faithfully the properties of water. The phase change energetics and the structures of the condensed phases of water-like toy models are not close to those of water, but the models provide insight on which microscopic interactions can produce water-like anomalous behavior. Examples of water-like toy models of water are the Mercedez-Benz model in two[15,16] and three dimensions[17], isotropic potentials with two characteristic length-scales[18-23] and modified van der Waals models.[24,25] All these models produce water-like anomalies and most of them also produce liquid-liquid transitions.

Existing coarse-grained models of water without electrostatics and hydrogen atoms represent intermolecular interactions with a spherically symmetric potential. It has been proved that isotropic potentials cannot reproduce the energetics and structure of water simultaneously.[14] Isotropic models that reproduce the radial distribution function (rdf) of liquid water are unable to reproduce the oxygen-oxygen-oxygen angular distribution function (adf),[14] they underestimate the internal energy of the liquid by about 50%,[13] and they do not produce the most characteristic anomaly of water, the existence of a density maximum.[14] Moreover, isotropic monatomic models of water do not form a tetrahedral crystal or a low-density glass on cooling. They model a "normal" liquid, not water.

To investigate whether a coarse-grained model can reproduce the structures and phase behavior of water without using electrostatics and hydrogen atoms, we first shift our attention from water to simple monatomic systems that also form tetrahedral structures: silicon and germanium. Similar to water, these elements form tetrahedral crystals at room pressure and have two amorphous phases:[26,27] a low- and a high-density glass. The low-density glasses - low-density amorphous ice (LDA), a-Si and a-Ge – are disordered structures with tetrahedral coordination.[26-28] The high-density glasses of these three substances also have analogous structure.[29]

The similarities between water, Si and Ge also encompass the phase diagram and anomalies. These three belong to a handful of substances whose liquid is denser than the crystal, resulting in a decrease of the melting temperature with pressure. The density of "normal" liquids increases monotonously on cooling. The density of water, on the other hand, displays a maximum at 4°C and sharply decreases in the supercooled region.[10] Silicon also displays a density maximum, deep in the supercooled regime.[30] The dynamics of these liquids are also anomalous: While the viscosity of "normal" liquids increases with pressure, liquid silicon and water become more fluid on compression.[10,31] This anomaly is more pronounced in the deeply supercooled regime, and disappears at higher temperatures.[10]

The similarities between these tetrahedral liquids suggest that water, as silicon, may be modeled as a single particle with only short-ranged interactions. This does not mean that electrostatic interactions or the hydrogen atoms are irrelevant in determining water structure and thermodynamics, but that their effect may be effectively produced with a monatomic short-ranged potential.

## II. Model and methods.

### A. The mW monatomic water model.

To "make water out of silicon", we start from the Stillinger-Weber (SW) silicon potential.[32] In the SW model, tetrahedral coordination of the atoms is favored by adding to a pairwise potential $v_2(r)$ a three-body term $v_3(r,\theta)$ that penalizes configurations with angles that are not tetrahedral, $v=v_2(r) + \lambda\, v_3(r,\theta)$. The parameter $\lambda$ tunes the strength of the tetrahedral penalty.[33] The higher the value of $\lambda$ the more tetrahedral the model.

The full expression of the SW potential as a function of the distances between pairs of atoms and the angles formed by triplets of atoms is given by

$$E = \sum_i \sum_{j>i} \phi_2(r_{ij}) + \sum_i \sum_{j \neq i} \sum_{k>j} \phi_3(r_{ij}, r_{ik}, \theta_{ijk})$$

$$\phi_2(r) = A\varepsilon \left[ B\left(\frac{\sigma}{r}\right)^p - \left(\frac{\sigma}{r}\right)^q \right] \exp\left(\frac{\sigma}{r - a\sigma}\right)$$

$$\phi_3(r,s,\theta) = \lambda\varepsilon [\cos\theta - \cos\theta_o]^2 \exp\left(\frac{\gamma\sigma}{r - a\sigma}\right) \exp\left(\frac{\gamma\sigma}{s - a\sigma}\right)$$

where[32] $A = 7.049556277$, $B = 0.6022245584$, $p = 4$  $q = 0$, and $\gamma = 1.2$ give the form and scale to the potential, the reduced cutoff $a = 1.8$ ensures that all terms in the potential and forces go to zero at a distance $a\sigma$, the cosine quadratic term around $\theta_o = 109.47°$ favors tetrahedral angles. The parameters $\lambda$ scales the repulsive three-body term and determines the strength of the tetrahedral interaction in the model; its value for silicon is 21.[32] Two additional parameters set the energy scale $\varepsilon$ (the depth of the two-body interaction potential) and the length scale $\sigma$ (the particle diameter) of the model. Note that the SW potential can be written in a reduced form independent of the values of $\sigma$ and $\varepsilon$. Only the tetrahedrality $\lambda$ and the size and energy scale, $\sigma$ and $\varepsilon$, are tuned to produce the monatomic water model mW that represents each molecule as a single atom with tetrahedral interactions.

### B. Simulation details.

We carried out Molecular Dynamics (MD) simulations using LAMMPS, a massively parallel MD software developed by Plimpton et al.[34] A reduced timestep of 0.025 was used for the parameterization and validation. For the final set of parameters we found that timesteps up to 10 fs (0.05 in reduced units) conserve the energy better than 1/10000 in microcanonical simulations of $10^6$ steps. We used a 10 fs step for the simulations to validate the properties of the model, except for the high pressure simulations and those that involve an open interface, where a 5 fs step was used. Where indicated, the temperature and pressure were controlled with the Nose-Hoover thermostat and barostat with relaxation times 1 and 2.5 ps, respectively. All isobaric simulations were at $p = 0$. Except when otherwise is indicated, the system contained 4096 particles in a periodic box and the simulation time was 10 ns.

### C. Property computation.

**Melting temperature.** The structures of hexagonal (*Ih*) and cubic (*Ic*) ices without hydrogen atoms correspond to hexagonal and cubic diamond, respectively. Their melting temperatures ($T_m$) were determined through the phase coexistence method, as implemented and discussed in detail in Ref. [35]. In this method, a perfect crystal and a liquid slab are put in contact to facilitate the growth of the stable phase on isobaric isothermal (NPT) MD run. Garcia Fernandez et al. applied this method to atomistic models of water and proved that it reproduces the melting temperatures obtained from free energy calculations.[35] We start from a periodic cell of dimensions approximately 50 Å x 30 Å x 30 Å, where half of it (~25 Å x 30 Å x 30 Å) is a perfect cubic or hexagonal diamond crystal and the other half is a liquid. In a NPT simulation starting from this system below $T_m$ ice grows until it encompasses all the system, same for the liquid above $T_m$. We determine $T_m$ as the mean value between the highest T for which it does not melt and the lowest for which it does, and report as error bar half the difference between these two. In the case of the mW potential, we estimated the precision from five independent series of simulations for each of the two crystalline structures.

**Density, enthalpy, heat capacity and compressibility.** The density was determined as $\rho = NM/(N_A \langle V \rangle)$, where where *N* is the number of particles in and *V* the volume of the simulation cell, $M = 18.015$ g is the molar weight of water, $N_A$ is Avogadro's number and $\langle ... \rangle$ indicates a time average over an equilibrium simulation. The enthalpies of the condensed phases were computed as $\langle H \rangle = \langle E + pV \rangle$, where *E* is the total energy per mol, *V* the simulation volume per mol and p the pressure of the system. We assumed the molar enthalpy of the vapor was that of an ideal gas with zero internal energy, $H_{gas}=1.5\ RT + pV_{gas}=2.5\ RT$. Relatively small systems, 512 or 576 particles, were used for the enthalpies calculation in the parameter search, while the results reported for the final mW potential were obtained with 4096 particles.

An isobaric quench simulation from 320 to 205 K in 230 ns (rate 0.5 Kns$^{-1}$) was used to compute i) the temperature dependence of the density and the location of its maximum, and ii) the enthalpy of the liquid, and its temperature derivative, *Cp*. The rate of change of the temperature is slow compared with the equilibration time of the liquid, and we assume that the liquid is in local equilibrium. *H(T)* and *ρ(T)* were computed from a rolling average over one nanosecond-length intervals. The assumption of local equilibrium was verified by computing the average density and enthalpy for 5 to 10 ns isothermal simulations at several temperatures along the whole temperature range; the average values are indistinguishable from those of the slow ramp. The use of the slow ramp is advantageous in determining the position of the density maximum without the need of interpolation. The enthalpy was fitted to an equation of the form[36] $H_{liquid}(T) = A + BT + C(T/T_o - 1)^\gamma$ (correlation coefficient 0.999824) from which the isobaric heat capacity was obtained by analytical derivation, $Cp = dH/dT)_{p,N}$.

The isothermal compressibility of mW liquid at 300K around $\rho_o = 1$ gcm$^{-3}$ was calculated by a finite difference approximation[37] as

$$\kappa_T = -\frac{1}{V}\left(\frac{\partial V}{\partial p}\right)_T \approx \left(\frac{\ln(\rho_2/\rho_1)}{p_2 - p_1}\right)_T$$

where $\rho_2$ and $\rho_1$ are 1% above and below $\rho_o$. The average pressures $p_2$ and $p_1$ were computed from a 10 ns NVT simulation at T = 300 K.

**Radial and angular distribution functions.** The pair distribution function between two water sites in the coarse-grained model was computed as an ensemble average over pairs of water particles

$$g(r) = \frac{V}{N^2}\left\langle\sum_{i=1}^{N}\sum_{j\neq i}\delta(r - r_{ij})\right\rangle.$$

The average number of neighbors in the liquid up to a distance $R_c$ is given by

$$n_c = 4\pi\frac{N}{V}\int_o^{R_c} r^2 g(r)dr.$$

The angular distribution function (adf) was computed as an ensemble average over the angles between each water and its closest $n_c$ neighbors

$$P(\theta) = \frac{1}{Nn_\theta}\left\langle\sum_{k=1}^{N}\sum_{i=1}^{n_c}\sum_{j\neq i}^{n_c-1}\delta(\theta - \theta_{ikj})\right\rangle$$

where $n_\theta$ is the number of angles subtended by the $n_c$ neighbors around the central molecule $k$. We selected $n_c = 8$ to compare with the neutron scattering results of Strassle et al.[38]

**Self-diffusion coefficient.** The diffusion coefficient of liquid water was computed from the slope of the mean square displacement with time using Einstein's relation

$$D = \lim_{t\to\infty}\frac{1}{6t}\langle|r(t) - r(0)|^2\rangle.$$

At room pressure, statistics were collected at temperatures ranging from 363 K to 243 K. To study the density dependence of the diffusion coefficient, NVT simulations were performed at 243 and 220 K at densities from 0.94 to 1.20 gcm$^{-3}$.

**Surface tension.** The liquid-vapor surface tension was determined as in Ref. [39]: a periodic liquid slab containing 1024 particles was placed between two empty regions, with its two interfaces perpendicular to the $z$ axis. The dimensions of the periodic cell containing the slab and the vacuum region is $Lx = Ly = 30$ Å and $Lz = 100$ Å. The surface tension was obtained from the average over 20 ns NVT simulation at 300 K of the components of the pressure tensors tangential and perpendicular to the liquid-vacuum interface, $\langle p_T\rangle$ and $\langle p_N\rangle$, respectively:[39]

$$\gamma = \frac{L_z}{2}\left[\langle p_N\rangle - \langle p_T\rangle\right].$$

The error was propagated from the uncertainties in $\langle p_T\rangle$ and $\langle p_N\rangle$.

**D. Parameterization of mW.**

To find the optimum values of $\lambda$, $\varepsilon$ and $\sigma$ we implemented a non-iterative procedure. First, we computed the melting temperature for $Ih$ in the range $22 < \lambda < 27$ in reduced units, $T_m^*(\lambda)$. Second, for each value of tetrahedral parameter $\lambda$, we found the energy scale $\varepsilon(\lambda)$ that yields the experimental $T_m$ of water: 273.15 K = $T_m^*(\lambda)\varepsilon(\lambda)/k_b$, where $k_b$ is Boltzmann's constant. Third, the phase change enthalpies were computed as a function of the λ, from which the value λ = 23.15 was selected as the one that best reproduce water's vaporization enthalpy (see Figure 1). Finally, the value of σ was scaled to reproduce the density of the liquid at 298 K.

The interaction parameters of mW are λ = 23.15, ε = 6.189 Kcal/mol and σ = 2.3925 Å; all other parameters are identical to silicon in Ref. [32] The potential is very short ranged: *all* forces between atoms farther than 4.32 Å are zero. The parameterization of mW places water as an element with tetrahedrality intermediate between silicon and carbon: Figure 2 shows that the tetrahedral strength of water, λ = 23.15 is higher than that of silicon λ = 21[32] and germanium λ = 20[26] and lower than that of carbon, λ = 26.2.[40] The tetrahedral ordering C > water > Si > Ge is supported by an increasing number of first neighbors in the liquids: carbon (<4)[41] < water (5.2-5.3)[42] < Si (~5.5-6)[43] < Ge(~6-7).[44]

We benchmarked mW against SPCE, the least expensive atomistic model, in simulations with 1600 molecules. mW is 180 times faster than SPCE. The speedup arises from the smaller number of particles (1 vs 3) the longer timesteps (10 vs 1.5 fs) and shorter range of interactions (cutoff at 4.32 Å vs Ewald sums).

### III. Results

**Energetics, density and surface tension.** The melting temperature $T_m$, enthalpy of sublimation of ice at $T_m$, enthalpy of vaporization of the liquid computed with molecular dynamics simulations of mW are within 2% of the experimental values, as shown in Table 1. In agreement with experiment, the mW model predicts that hexagonal ice ($T_m$=274.6 ±1 K) is more stable than cubic ice ($T_m$=271.5 ± 1 K). The density of mW liquid is within 1% of the experimental value in the temperature range 250 to 350K.

How well a water model reproduces the liquid-vapor surface tension is of the highest relevance for the study of water at the vacuum and hydrophobic interfaces, wetting-drying transitions and hydrophobic attraction. The liquid-vacuum surface tension of mW at 300K is $\gamma_{lv}$ = 66 ± 2 mJ/m$^2$, in excellent agreement with the experimental value, 71.6 mJ/m$^2$.

**Structure.** Simple liquids, such as molten metals, typically have an average of ~11 first neighbors. At 25 °C, water has an average of 5.1 to 5.3 molecules in the first coordination shell and characteristically short-ranged radial ordering.[42,45] The radial distribution function was *not* considered in the parameterization of mW. Nevertheless, Figure 3 shows that the structure of the mW liquid is in excellent agreement with the one derived from X-ray and neutron diffraction experiments for water.[45][42] (Figure 3b). The number of water neighbors up to a distance of 3.5 Å is between 5.1 and 5.3 for the neutron/X-ray refined structures of Soper[42] and is 5.1 for the monatomic water (mW has 4.25 neighbors within the first 3.3 Å).

The angular distribution function (adf) provides a more stringent validation for the quality of a water model. The monatomic model quantitatively reproduces the experimental OOO adf of liquid water[38] (Figure 3a). The intermolecular forces in the mW model vanish at just 4.3 Å, so we conclude that long-range forces are not needed to reproduce the characteristically short-ranged structure of liquid.

**Density Anomaly.** Among water thermodynamic anomalies, the best known is the density maximum at 4°C. Most atomistic models of water reproduce the existence of a density maximum with varied success in predicting the temperature of maximum density (TMD). Figure 4 shows the liquid density as a function of temperature at room pressure for water, mW and atomistic models; Table 1 summarizes the TMD and maximum densities, $\rho_{liquid,MAX}$. The density maximum of mW is 1.003 gcm$^{-3}$, in excellent agreement with the experimental value, 0.99997 gcm$^{-3}$.[46] The temperature of maximum density (TMD) of mW is 250 K, below the melting temperature and the experimental value of 277 K.[46] While the TMD is an intrinsic property of the liquid, the melting point depends on the relative enthalpy and entropy of liquid and crystal. mW was parameterized to reproduce the experimental melting temperature, but it can be argued that monatomic water should have a melting point higher than molecular water, because there is no contribution from the rotational entropy to the melting of the monatomic liquid. We interpret that the location of the TMD in mW below $T_m$ (as also observed in silicon[30]) is a consequence of the monatomic character of the model.

**Heat Capacity Anomaly.** Another consequence of mW being monatomic is a low heat capacity. mW has one third of the degrees of freedom of atomistic water, and a constant pressure heat capacity $Cp$ at 25°C that is 44% of the experimental value (33 vs 75.3 J/Kmol[46]). The low value should be mainly due to the lost of the rotational contribution to the liquid's heat capacity.

In Figure 5 we present the heat capacity of liquid water and mW, with respect to their values at 300K. There is a sharp increase in the $Cp$ of supercooled liquid water,[47] that correlates with the dramatic volume expansion shown in Figure 4. The coarse-grained model mW reproduces this thermodynamic anomaly associated to the transformation of the liquid to a low-density almost perfectly tetrahedral amorphous phase (see below). The experimental heat capacity, available down to 245K, is well represented by $C_p(T)=0.44(T/222-1)^{-1.5}+74.3$.[36] The heat capacity of monatomic water mW is well described by $C_p(T)=2.36 (T/185-1)^{-1.5}+28.25$ in the temperature range 205 to 320 K. The temperature of the transformation is shifted to lower temperatures with respect to the experiment for the same reasons discussed above for the density maximum.

**Diffusion Anomaly.** The diffusion coefficient of mW at 298K is D=6.5 $10^{-5}$cm$^2$/s, almost three times the experimental value (see Table 1). The mobility in mW is faster because the molecules are not slowed down by the reorientation of hydrogen atoms. The effect of the lack of hydrogens is not only an increase in the magnitude of the mobility but also a lower activation energy than the experiment: Figure 6 shows that D of mW is less sensitive to temperature than that of the experimental substance. The result is that mW reaches the deeply supercooled state where the liquid transforms to a low-density structure, with relatively high mobility.

Experimentally and in atomistic simulations with the SPC/E model, water diffusivity attains a maximum when the liquid is compressed to a density of about 1.1 gcm$^{-3}$.[48,49] The coarse-grained model reproduces this anomalous density dependence: the diffusivity passes through a maximum for a density of 1.1 and 1.08 gcm$^{-3}$ at T=243 and 220K, respectively (Figure 7). The ratio $D_{max}/D(\rho=1gcm^{-3})$ –the strength of the anomaly- is comparable in the experiment and coarse-grained simulations if the temperature is measured from the TMD: at 25K below the TMD, the enhancement in diffusivity is 1.8[50] and 1.75, respectively.

**Phase transformations of supercooled water.** The existence of a density maximum and a heat capacity that dramatically increases in supercooled liquid silicon[33] and water is related to the stabilization of low-density amorphous structures (a-Si and LDA) at low temperatures. Computer simulations of Si with the Stillinger-Weber potential reveal a first order liquid-liquid transition at room pressure.[51] It is still debated whether –and in which pressure range- a first order transition separates the high- and low-density liquids in water.[28] On the one hand, experimental studies are hindered by the crystallization of the metastable liquid when it approaches the putative location of the liquid-liquid coexistence line. The easy crystallization makes it difficult to study the characteristics of deeply supercooled water and the process of vitrification or ice nucleation in experiments. On the other hand, the slow dynamics of the supercooled liquid hinders its study through atomistic simulations. The monatomic model, with its low computational cost and higher mobility, is adequate to fill in this gap in the study of phase transitions and properties of supercooled water.

As observed in the experiments, the product – ice or glass- of a fast quenching of the monatomic liquid water at room pressure depends on the cooling rate: we find that mW forms ice for cooling rates $10^9$K/s or slower. At higher quenching rates, mW water transforms to a low-density liquid (LDL) that vitrifies to LDA (see Figure 8). It is interesting to note that crystallization in the quenching simulations happens always around the temperature where the high-density liquid transforms into the low-density one, $T_{LL}$= 202 K for the mW model at 1 atm. More studies are needed to determine whether the liquid-liquid transformation is continuous or first order.

The cooling rate needed to bypass crystallization in a system with 4096 mW is $\sim 10^3$ faster than in experiments involving micron-sized droplets: ice nucleation in mW is several orders of magnitude faster than in real water. The reasons are probably twofold: i) the lack of hydrogens that reduce the search in configurational space to produce ice nuclei, and ii) the higher diffusivity of the liquid, also due to a lack of hydrogen atoms. The highest rate makes feasible the collection of the thousands of crystallization trajectories needed to characterize the stochastic process of ice nucleation. It should be noted that the ice nucleation times are a strongly varying function of the temperature and a system with 4096 mW can be equilibrated down to 205 K without interference of crystallization: In this condition, the characteristic time for ice nucleation is 30 ns while the relaxation time of the monatomic liquid is less than one nanosecond.[52] The study of the mechanism of ice nucleation in bulk and in nanopores and its relationship to water polyamorphism will be presented in separate communications.

If crystallization is bypassed it is possible –but difficult!- to partially relax the low density liquid at a temperature below $T_{LL}$. The relaxed density for a system of 512 molecules after 130 ns NPT simulation at 190 K is shown as a cross in Figure 8. The structure of mW's LDL is an amorphous tetrahedral network with an average of 4.04 first neighbors and rdf in excellent agreement with the one for LDA measured by neutron diffraction (Figure 3c). The formation of amorphous ice, not considered in the parameterization of mW, supports the hypothesis that a monatomic model with short-ranged tetrahedral interactions is enough the produce the main features of water's phase behavior at room pressure.

**IV. Discussion.**

Can a coarse-grained model without electrostatic interactions and hydrogen atoms reproduce water properties as accurately as all-atoms models? Table 1 compares the performance of mW, SPC, SPCE, TIP3P, TIP4P and TIP5P in representing key properties of water at room temperature and the melting point. mW outperforms the atomistic models in six out of the ten properties listed in the table: the prediction of hexagonal ice as the stable crystal at room pressure and its melting point, the enthalpy of melting of ice, the density of the liquid at $T_m$ and 298K, the maximum density of the liquid and the liquid-vapor surface tension. Of the other four, the enthalpy of vaporization is just 1.2% above the experimental value for mW. The enthalpy of sublimation of ice (not reported for most atomistic models) is only 1.7% higher than experiment. Let's address now the three properties for which mW is outperformed by at least one atomistic model: The predicted temperature of maximum density, 27K below experiment, is in the middle of atomistic range (worst: TIP3P, 95K below; best: TIP5P, 8K above). The diffusion coefficient is the only property of Table 1 for which mW trails all atomistic models: mW predicts a value 2.8 times the experiment, while atomistic models predict from 1.04 (best, SPCE) to 2.3 (worst, TIP3P) of water's value. The second worse reproduced property is the density of ice, overestimated by all models, for which mW is better only to TIP5P (best: SPC).

Overall, mW outperforms the most popular atomistic models in the representation of the ten properties of Table 1. But there is a price paid for the lack of hydrogens: one the one hand, mW cannot "extend and bend" hydrogen bonds as water does, resulting in i) a reduced density gap between liquid water and ice, and ii) a lower isothermal compressibility, $\kappa_T \approx 1.9 \; 10^{-5} \text{atm}^{-1}$ at 300K compared with the experimental value of $4.58 \; 10^{-5} \text{atm}^{-1}$:[46] While it may be possible to improve the flexibility of the model to better reproduce the compressibility and ice density without significant deterioration of other properties, it is not clear to us that this can be done while keeping a simple form of the intermolecular interactions. One the other hand, the lack of hydrogen atoms is responsible for the highest diffusivity of the monatomic model: coarse-grained models evolve on smoother potential energy surfaces than fully atomistic ones,[12] and the hydrogen's effectively produce a friction on water's center of mass translation.

The true Achilles heel of coarse-grained models is the heat capacity: a model with less degrees of freedom necessarily underestimates $Cp$. Water's rotational contributions to the heat capacity – active in the liquid and vapor phases- are absent in the monatomic model. The underestimation of $Cp$ will produce a degradation of the agreement in the energies and entropies as the temperature moves away from the one used in the parameterization (273 and 298K, in this case).

mW displays the diffusional and thermodynamic anomalies of water. We note that the density of maximum diffusivity and the magnitude of the enhancement are in very good agreement with the experiment, although the pressure is overestimated, due to a low compressibility. This supports a structural origin for the diffusivity maximum in water. It would be interesting to determine whether mW reproduces the hierarchy of anomalies[53] observed for atomistic models of water. The thermodynamic anomalies are produced by a sharp high- to low- density transformation of the liquid at $T_{LL}$ that is fifty degrees below the TMD, as observed in experiments of nanoconfined water.[54] We computed the heat capacity of the liquid down to a few kelvins above $T_{LL}$ and found a power law behavior (Figure 3) that predicts divergence at a temperature 17 K below the actual $T_{LL}$ of the model. These results, and the observation of ice nucleation from supercooled water suggest that mW will be useful in understanding the puzzling behavior of water at low temperatures, close and inside "no man's land".[28]

The monatomic tetrahedral model faithfully reproduces the structure of ice, liquid water and low-density amorphous ice using extremely short ranged interactions: all forces go smoothly to zero at 4.32 Å, a distance shorter than the second peak in the liquid's rdf; compare this with the long ranged electrostatic forces used in atomistic simulations of water. We conclude that long-ranged interactions are not needed to model the structure of water. The introduction of a nonbond angle dependent term in the coarse-grained interaction potential is essential to capture the physics of water intermolecular interactions, and results in a model of water in which the molecules are "hydrogen bonded" although there are no hydrogen atoms. The hydrogen atoms can be regarded as the "glue" that keeps the oxygens in hydrogen-bonded positions.

How well can the monatomic water model reproduce the structure and properties of aqueous solutions and water at interfaces? Electrostatic interactions are essential for the solvation of ions and hydrophilic molecules, but mW does not speak the language of electrostatics. It is necessary to mimic the effect of these interactions through short-ranged potentials to preserve the computational efficiency of the coarse-grained model. Even if the efficiency was not a concern, the use of electrostatics for the solute-solute interactions does not address the problem of how do water and solute interact without electrostatics. Preliminary results from our group show that it is possible to reproduce the main effect of hydrophilic and ionic solutes on the structure of water, the decrease in tetrahedrality evidenced in the experiments by the depression of the second peak in the OO rdf,[55] with ony short-ranged SW potentials.[56] It is still an open question whether this can be extended to model two challenging properties of ionic solutions: the stabilization of solvent separated ion pairs in aqueous solutions and the layering of cations and anions at different depths from the water-vacuum interface.[57]

It has been reported that a good description of hydrophobic effects in simulations correlates with an accurate description of the liquid density over a broad temperature range.[58] The signatures of the hydrophobic effect have been traced to water's low compressibility and relatively low decrease of density on heating, compared to organic solvents.[59] Recently, Buldyrev et al.[23] found that the Jagla model -an isotropic ramp potential with two characteristic length-scales that displays the thermodynamic, structural and diffusional anomalies as water but not water's characteristic liquid and crystal structures- produces water-like solvation thermodynamics for hydrophobic solutes: a solubility minimum as a function of temperature and swelling of hydrophobic polymer chains at low temperature. Their study suggests that water-like solvation of hydrophobic molecules may be given by the ability of the solvent to expand on cooling. The density of liquid mW is within 1% of experiment for 250 < T < 350 K; in better agreement than the atomistic models (Figure 3 and Table 1) in spite of the low TMD. The extent by which mW can predict hydrophobic hydration remains to be studied, but the good agreement in the density and its temperature dependence, energetics, structure and surface tension suggests that mW will be a realistic water solvent for hydrophobic molecules in coarse-grained simulations.

An interesting question is whether the monatomic model, parameterized from bulk data, can reproduce interfacial properties of water. We have shown above that mW reproduces the liquid-vapor surface tension of water at ambient conditions. In work to be reported elsewhere,[60,61] we found that the monatomic model produces the phase behavior of interfacial atomistic models of water in hydrophobic confinement: mW confined between nanoscopic hydrophobic disks displays wetting-drying transitions[60] at surface separations in good agreement with those found in atomistic studies[62,63] and –at lower temperatures- it forms bilayer ice and other ice structures related to bulk hexagonal ice [61], also observed in atomistic simulations.[64,65]

### V. Conclusions.

Tetrahedrality, through the formation of hydrogen bonds, is arguably the defining characteristic of water interactions. Head-Gordon and Rick found that modified SPC/E and TIP4P-Ew models that form only two hydrogen bonds, do not produce water-like properties.[66] Debenedetti and coworkers reached the same conclusion for SPC/E potentials for which the H-O-H angle is modified to hinder the tetrahedral coordination of the molecules.[67] In this work, we strip water of atomistic detail and represent it as an atom with very short-ranged tetrahedral interactions. The success of the mW model in reproducing the liquid, crystal and glass structures of water, their energetics, liquid anomalies, and the corresponding phase transitions strongly indicates that the nature of the intermolecular interactions –covalent/metallic or dipole/hydrogen bond- is less defining of the structural and thermodynamic behavior of these substances than the formation of tetrahedral configurations. More provocative, the monatomic water model mW is just a more tetrahedral silicon atom, with the corresponding change in energy and density scale. Only one of the seven parameters of the reduced Stillinger-Weber potential for silicon is tuned to produce a model that is surprisingly accurate in the description of water. Water and silicon not only belong to the same family: they are close siblings.

Angell et al. have qualitative positioned water within the family of tetrahedral liquids and conclude that water behavior is intermediate between silicon and silica.[68] In developing mW we move a step further and quantify how different water, silicon carbon and germanium are, in terms of a single parameter: the strength of the tetrahedral interactions. This quantitative relationship provides a unified framework to understand the rise and death of anomalous behavior along the family of tetrahedral liquids. Results in this respect will be presented in a future communication.

mW is a model without hydrogens and electrostatics but, of course, there are properties of water that *require* the electrostatics and the hydrogen atoms for their description (e.g. dielectric properties, rotational dynamics, all its chemistry!). The mW coarse-grained model does not replace atomistic representations of water but provides insight on which intermolecular interactions are responsible for water behavior. We conclude that the lack of hydrogen's has more impact on water properties –lower heat capacity, lower structural flexibility to accommodate compression, less hindered diffusivity - than the shortening of the intermolecular interactions to 4.32 Å. There is an increasing interest in developing theories[69] and models[70] to replace the long-ranged electrostatic interactions by effective short range potentials in all atom simulations. Izvekov et al. recently coarse-grained the interactions of SPC and TIP3P models to produce fully atomistic models where the electrostatic interactions are replaced by a function that vanishes at 10 Å.[70] The atomistic short-ranged potentials reproduce the rdf, density, internal energy, compressibility and diffusion coefficient of the original models. The success of this "coarse-graining in interaction space"[70] supports our conclusion that the topology of the interactions, and not the range of the potential, is the key to model water.

The most severe representability issue of isotropic monatomic water models, namely their inability to simultaneously reproduce the structure and energetics of water at any state point, is removed by the introduction of the tetrahedral interactions. The monatomic tetrahedral model predicts the studied water properties - with the notorious exception of the response functions- with comparable or better accuracy than atomistic models. The use of anisotropic interactions does not degrade the efficiency of the model: mW is two orders of magnitude faster than the least expensive atomistic model.

Coarse-grained models of polymers, proteins, carbohydrates, biomembranes, etc, have been developed in recent years. We expect that mW will be combined with these or new models to produce a computationally efficient representation of water in coarse-grained simulations of biomolecules and materials. The accuracy of coarse-grained models in reproducing the properties of solutions and interfacial water –particularly for ion containing systems- is a question that deserves further study.

**Acknowledgements:** This research was supported by NSF under Collaborative Research Grant CHE-0628257. We acknowledge the Center of High Performance Computing at the University of Utah for a generous allocation of computing time. We thank Alan Soper and Thierry Strassle for sharing their data on the radial and angular distribution functions of water, respectively; and Jack Simons and Austen Angell for their comments on an earlier version of the manuscript.

**Figure 1.**

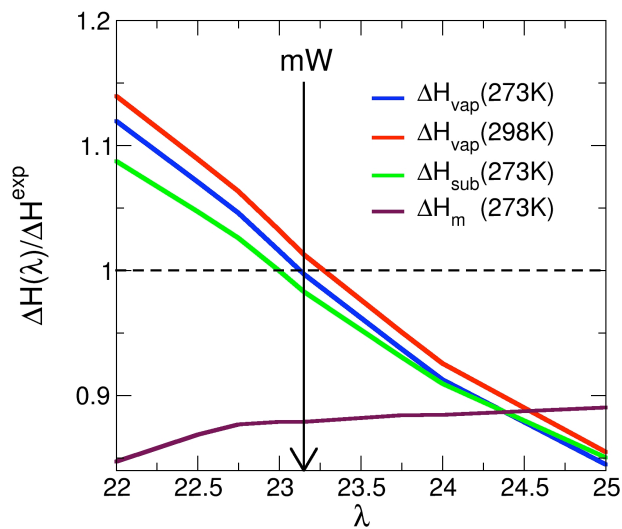

**Figure 1.** (color online) Optimization of the tetrahedral parameter λ for the monatomic water. The ratio between the enthalpies of vaporization, sublimation and melting in SW potentials and the experiment shows best agreement for a tetrahedrality λ = 23.15. The energy scale for each of these potentials is obtained by requiring that the computed melting point agrees with the experimental values for hexagonal ice.

**Figure 2.**

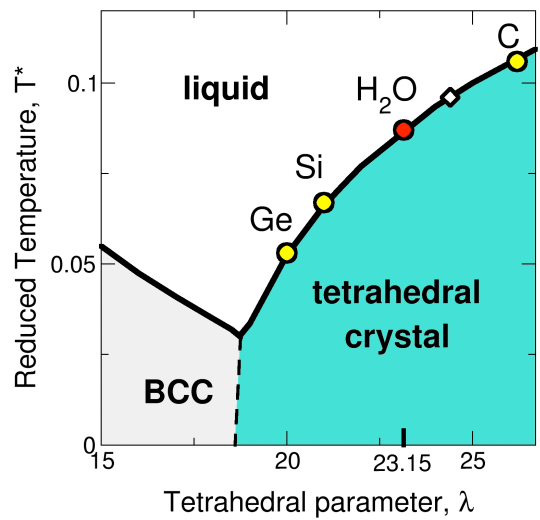

**Figure 2.** (color online) Phase diagram of modified SW potential as a function of the strength of the tetrahedral repulsive parameter $\lambda$, at zero pressure. The stable crystal is tetrahedral for $\lambda >$ 18.75; for less tetrahedral potentials an 8-coordinated BCC crystal is more stable.[33] Carbon, water, silicon and germanium can be considered as members of this family with different tetrahedral strength: $\lambda_C= 26.2$,[40] $\lambda_{water}= 23.15$ (this work), $\lambda_{Si}=21$,[32] and $\lambda_{Ge}=20$.[26] Their reduced melting points, $T_m k_b/\varepsilon$, are indicated by circles on the coexistence curve. The hollow rhomboid signals the tetrahedrality ($\lambda= 24.4$ at p=0) for which the coexisting crystal and liquid have the same density. Carbon, with $\lambda > 24.4$, is the only one of these substances for which the crystal (both diamond and the most stable graphite) is denser than the liquid.[41]

**Figure 3.**

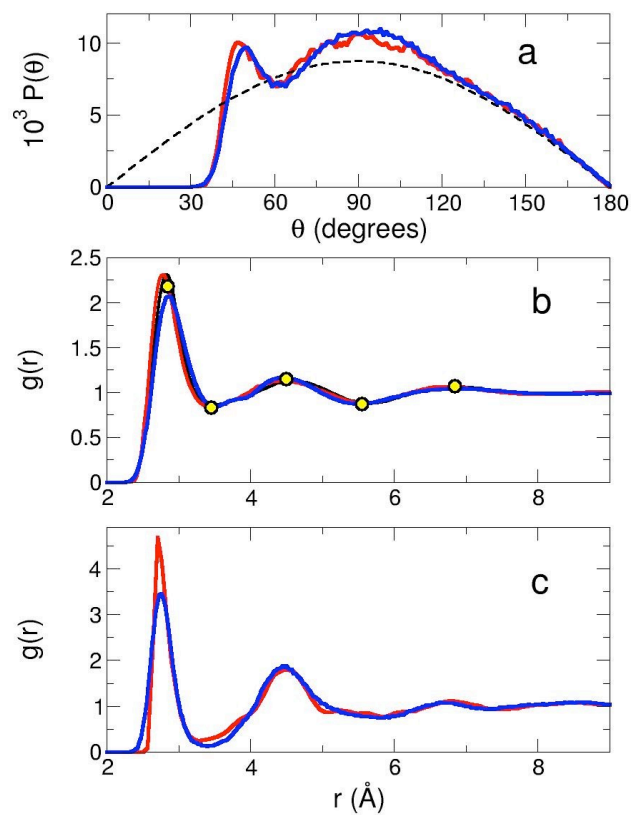

**Figure 3.** (color online) The tetrahedral monatomic model without electrostatic interactions reproduces the structure of water. a) Angular distribution function of eight closest oxygen neighbors in water at 298K in experiment[38] (red) and mW simulation (blue). Dashed line is the random distribution. b) Radial distribution function of liquid water at 298K in mW (blue line) and experiment: X-ray diffraction from Ref. [45] (yellow circles) and refined structure from Advanced Light Source Xray and neutron data, from Ref. [42] (red and black lines). c) The experimental radial structure of LDA[77] (red) is well reproduced by the mW model (blue).

**Figure 4.**

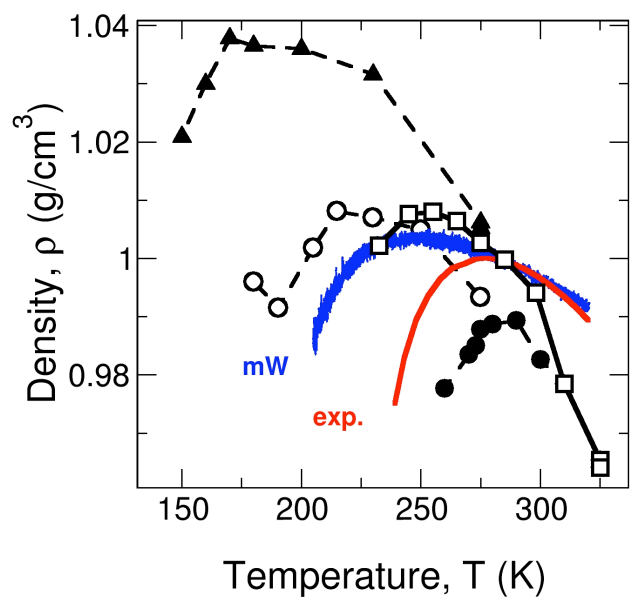

**Figure 4.** (color online) Temperature dependence of the density of liquid water at p=1 atm. The experimental (labeled exp) density maximum is qualitatively reproduced by all atomistic models of water and the monatomic model with tetrahedral interactions mW, but not by isotropic pair potentials that reproduce the radial distribution function of water.[14] Atomistic data from Ref. [76]: TIP5P (black circles), TIP4P (white squares), TIP3P (black triangles), SPC (white circles). Experimental density from Ref. [46].

**Figure 5.**

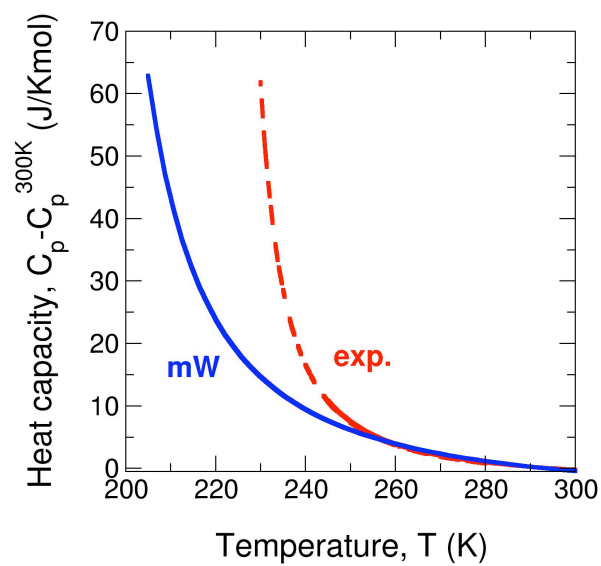

**Figure 5.** (color online) The constant pressure heat capacity of liquid water shows a marked increase in the supercooled region, coincident with the expansion of the density (Figure 4). Experimental data, available down to 245K, is well represented by $C_p(T)=0.44(T/222-1)^{-1.5}+74.3$.[36] The dotted line extrapolates the fit into the temperature range experimentally inaccessible due to ice crystallization. The heat capacity of monatomic water mW is well described by $C_p(T)=2.36 (T/185-1)^{-1.5}+28.25$ in the temperature range 205 to 320K.

**Figure 6.**

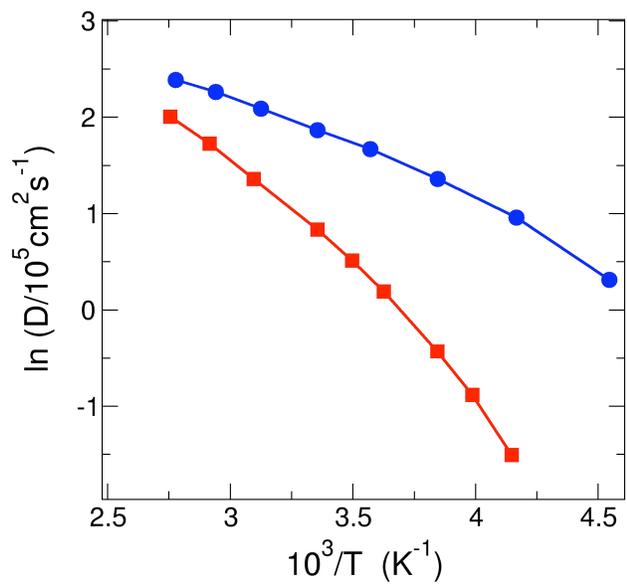

**Figure 6.** (color online) Diffusion coefficient of mW and experimental water as a function of temperature. The diffusion coefficient of monatomic water (blue circles) is higher and less sensitive to temperature than the experimental one (red diamonds).

**Figure 7.**

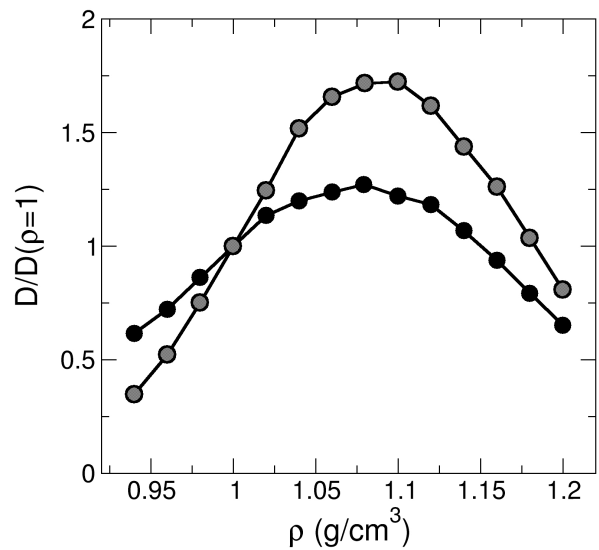

**Figure 7.** The monatomic water reproduces water's diffusivity anomaly. Relative diffusion with respect to that at $\rho = 1$ g/cm$^3$ at 243 K (black circles) and 220 K (gray circles).

**Figure 8.**

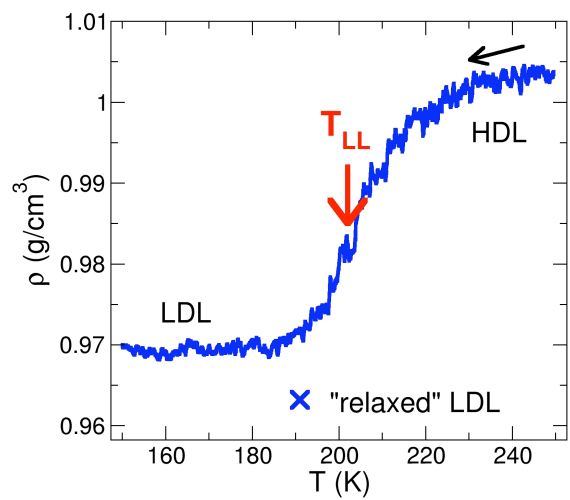

**Figure 8.** (color online) Liquid-liquid transformation in supercooled liquid water. The density of liquid water through a linear temperature quench at a 10K/ns rate displays a sharp transition at $T_{LL}$ =202 K from a high-density structure (HDL) to a low-density one (LDL). Relaxation of the liquid below $T_{LL}$ produces a liquid of lower density, indicated with a red cross. The liquid-liquid transformation competes with ice crystallization, that occurs around $T_{LL}$ for quenching rates 1K/ns.

**Table 1.** Comparison of water models and experiment.

| | $T_m$ HEX. ICE (K) | $\Delta H_m$ ($T_m$) (kcalmol$^{-1}$) | $\rho_{liquid}$ ($T_m$) (gcm$^{-3}$) | $\rho_{ice}$ ($T_m$) (gcm$^{-3}$) | $\rho_{liquid}$ (298K) (gcm$^{-3}$) | $\Delta H_{vap}$ (298K) (kcal/mol) | $D$ (298K) ($10^{-5}$cm$^2$s$^{-1}$) | $\gamma_{LV}$ (300K) mJm$^{-2}$ | TMD (K) | $\rho_{liquid, MAX}$ (TMD) (gcm$^{-3}$) |
|---|---|---|---|---|---|---|---|---|---|---|
| *Exp.* | *273.15* | *1.436* | *0.999* | *0.917* | *0.997* | *10.52* | *2.3* | *71.6* | *277* | *0.99997* |
| mW | **274.6** | **1.26** | **1.001** | 0.978 | **0.997** | 10.65 | 6.5 | **66.0** | 250 | **1.003** |
| SPC | (191) | 0.62 | 0.991 | **0.934** | 0.977 | **10.56** | 4.0 | 53.4 | 228 | 1.008 |
| SPCE | (215) | 0.74 | 1.007 | 0.950 | 0.999 | 10.76 | **2.4** | 61.3 | 241 | 1.012 |
| TIP3P | (146) | 0.30 | 1.017 | 0.947 | 0.986 | 10.17 | 5.3 | 49.5 | 182 | 1.038 |
| TIP4P | 232 | 1.05 | 1.002 | 0.940 | 1.001 | 10.65 | 3.9 | 54.7 | 253 | 1.008 |
| TIP5P | (274) | 1.75 | 0.987 | 0.982 | 0.999 | 10.46 | 2.6 | 52.3 | **285** | 0.989 |

**Table Caption:**

Melting temperatures of hexagonal ice, densities of liquid and crystal phase at coexistence and enthalpy of melting from Ref. [71]. Parentheses enclosing a $T_m$ signal that the stable crystal is ice II, not hexagonal ice, for these models.[72] Diffusion coefficients $D$ and density at 298K from Ref. [73] and [74]. Liquid-vacuum surface tensions from Ref. [75]. *TMD* and its corresponding liquid density $\rho_{liquid,MAX}$ from Ref. [76]. Bold numbers signal the closest agreement with the experiment.